%% file: _main.tex
\documentclass{article} % For LaTeX2e
\usepackage{iclr2026_conference,times}

\input{math_commands.tex}

\usepackage{hyperref}
\usepackage{cleveref}
\usepackage{url}
\usepackage{graphicx} 
\definecolor{darkblue}{rgb}{0,0, 0.7}

\hypersetup{colorlinks=true, citecolor=darkblue, linkcolor=darkblue, urlcolor=darkblue}

\title{The Invisibility Hypothesis: Promises of AGI and the Future of the Global South}

\author{L. Julián Lechuga López\\
Tandon School of Engineering, New York University\\
New York University Abu Dhabi\\
\texttt{ljl5178@nyu.edu} \\
\And
Luis Lara \\
Mila, Montreal, Quebec, Canada \\
\texttt{luis.lara@mila.quebec} \\
}

\iclrfinalcopy % Uncomment for camera-ready version, but NOT for submission.
\begin{document}

\maketitle

\input{0_abstract}

\input{1_introduction}

\input{2_promises}

\input{3_futures}
\input{conclusion}

\clearpage
\bibliography{references}
\bibliographystyle{iclr2026_conference}

\end{document}

%% file: math_commands.tex
\usepackage{amsmath,amsfonts,bm}

\def\eqref#1{equation~\ref{#1}}
\def\1{\bm{1}}

\DeclareMathAlphabet{\mathsfit}{\encodingdefault}{\sfdefault}{m}{sl}
\SetMathAlphabet{\mathsfit}{bold}{\encodingdefault}{\sfdefault}{bx}{n}

%% file: 0_abstract.tex
\begin{abstract}
Discussions surrounding Artificial General Intelligence have largely focused on technical feasibility, timelines, and existential risk, often treating its social impact as being the same across different populations. 
Less attention has been paid to how advanced AI systems may interact with existing global inequalities. 
This paper examines the implications of AGI for people in the Global South, arguing that the availability of highly autonomous, general-purpose cognitive systems does not guarantee equitable outcomes.
We establish that, as scientific discovery, economic coordination, and governance become increasingly automated, the relevance of human individuals may become conditional on access to infrastructure, institutional inclusion, and geopolitical circumstances rather than skills or intelligence. 
Under this setting, the Global South faces different pathways: in the best case, geographic location is no longer relevant as AGI fully democratizes access to knowledge and essential services for everyone in the globe; in the worst case, existing structural constraints are severely amplified, rendering already marginalized populations not merely economically invisible, but functionally irrelevant to global systems. 
We ground this analysis in empirical signals from contemporary AI deployment and extend to potential trajectories, highlighting both risk and opportunity pathways for Latin America, Africa, and South Asia.
\end{abstract}

%% file: 1_introduction.tex
\section{Introduction}

The significant advances in large-scale AI systems over the past five years have intensified debates around Artificial General Intelligence (AGI), shifting attention from whether such systems are deemed possible to when they might emerge and what their implications may be \citep{amodei_machines_of_loving_grace, bikkasani2025navigating}.
Whereas explicit discussion of AGI was until recently viewed as speculative or even taboo within academic research, the conversation has largely moved from “will we ever see these systems?” to “when will they inevitably appear?”\citep{ openai_planning_for_agi, salesforce_agi, deepmind_responsible_agi}.
While technical discussions dominate the conversation, questions concerning the social, political, economic, and technological consequences of advanced AI systems for developing countries remain comparatively underexplored \citep{wall2021artificial}.
This becomes particularly relevant for countries commonly grouped under the umbrella of Global South. 
The term itself is relatively recent and often functions less as a precise geographical category and more as a morally sanitized label for regions that have historically occupied disadvantaged positions in the global political economy \citep{dados2012global}. 
Across these regions, patterns of technological adoption and socioeconomic development have frequently reinforced dependency rather than enabling sustained, autonomous growth \citep{dirlik2007global}.

In this context, AGI should not be viewed solely as a future technological transformation, but as a conceptual endpoint that will stress-test existing global disparities. 
Rather than removing differences across societies, increasingly capable AI systems may generate very different trajectories between countries that already occupy strong positions of economic, institutional, and geopolitical power \citep{korinek2024scenarios, restrepo2025we}. 
The effects discussed in this paper are therefore not limited on the sudden arrival of AGI, but are unfortunately already visible in contemporary AI systems and likely to intensify as they become more capable, autonomous, and economically central.

Although we operate under the hypothetical assumption that AGI may eventually be achieved, this paper adopts a cautious and agnostic stance regarding claims of its imminent arrival. 
% Instead, we focus on observable trends in existing AI systems and extrapolate toward more advanced intelligent systems. 
Therefore, we ask a simple question: As scientific discovery, economic policies, and government strategies become increasingly automated, what do this imply for populations that are poorly represented in data, weakly integrated into formal economies, and distant from global centers of technological power?

\section{The Invisibility Hypothesis}

This paper advances the \emph{Invisibility Hypothesis}, which describes a specific failure mode of AI-controlled economic and political management. 
Here, invisibility is not used metaphorically. 
We employ the term operationally to denote the absence of actionable representation within systems increasingly controlled by AI \citep{hao2025beyond}.

An individual is economically invisible when their identity, transactions, output quality, or risk cannot be reliably observed, standardized, or audited by the pipelines that govern credit, procurement, insurance, logistics, and policy priorities \citep{potot2008strategies, hatton2017mechanisms}. 
Conversely, we refer to individuals as machine-legible when they possess stable identities, standardized and traceable transactions, and outputs that can be measured and compared within data-rich infrastructures \citep{he2024digital}. 
This distinction matters because AI-mediated allocation tends to optimize for what is measurable and verifiable, not necessarily for what is socially valuable.
Hence, we propose the following central idea:

\begin{quote}
\textbf{Invisibility Hypothesis.} As AI systems approach advanced general-purpose capability and increasingly function as the coordination layer for economic and political allocation, decision-making process will work under the constraints of digital verification, audit, and standardization of processes. 
As a result, these systems systematically favor machine-legible individuals rendering large segments of the Global South, particularly informal workers and small-scale producers, economically invisible through managed exclusion rather than direct displacement.
\end{quote}

Crucially, for these "invisible individuals", the short-term risk is not replacement but the loss of access and relevance within AI-controlled credit systems, supply chains, insurance markets, and policy processes. 
Once exclusion begins, it can become self-reinforcing: being denied credit, contracts, insurance coverage, or program eligibility reducing the production of verifiable records and standardized signals, further lowering visibility and making reentry increasingly difficult \citep{hao2025empire}.

In this view, routine office work in highly-digitized environments may be displaced early because they are already standardized into the interface that the AI understands. 
In contrast, many physical jobs may only be displaced if reliable, low-cost robotics actually matures. 
The distinctive harm captured by this hypothesis, however, is that exclusion via AI-controlled geoeconomics can precede full scale automation entirely, quietly shrinking participation and relevance even when the underlying work remains economically or socially valuable.

%% file: 2_promises.tex
\section{The Promises of AGI}

The current ideas of AGI frame it as a transformative force capable of changing long-standing constraints on human progress \citep{amodei_machines_of_loving_grace, hao2025empire,doker2025wicked}. 
Rather than treating AGI as a discrete technical milestone, we treat it here as a holistic test for global economic and political structures. 
In this view, AGI is not a binary event, but a total shift in which highly autonomous systems perform a substantial fraction of scientific discovery, economic coordination, and administrative decision-making.
Under this framing, the central question is not whether such systems are technically achievable, but if the promises commonly associated with it (e.g., self improvement, efficient coordination, fully optimized problem-solving) actually translate into equitable outcomes across societies with radically different material and institutional conditions.

\paragraph{Capability Without Omnipotence.}

Public discussions around AGI often oscillate between two extremes: systems portrayed either as narrowly useful tools or as near-omnipotent entities. 
In practice, even highly capable general-purpose systems remain constrained by physical resources, infrastructure, governments, and geography. 
Intelligence alone does not immediately eliminate bottlenecks related to energy availability, compute access, data quality, institutional trust, or political authority \citep{amodei_machines_of_loving_grace}.
For many economic and social problems, additional intelligence yields diminishing returns once coordination, infrastructure, and enforcement become the dominant constraints \citep{amodei_adolescence_of_technology}. 
A system may be capable of designing optimal agricultural policies, medical interventions, or efficient supply chains, yet remain ineffective if surrounding institutions cannot implement, regulate, or absorb these recommendations. 
In this sense, AGI does not abolish structural inequality, it interacts with, and is shaped by the existing structures of human power and capacity.

\paragraph{Global South.}

We consider this term not as a moral category, but as shorthand for regions that occupy structurally disadvantaged positions in the global political economy. 
These regions typically combine high population share with lower GDP per capita, elevated informality, weaker and often corrupt institutions, and limited control over strategic technologies and infrastructure (\Cref{fig:futures}) \citep{dirlik2007global, dados2012global}.
Historically, such conditions have shaped industrialization and development pathways by constraining access to capital, expertise, and markets. 
If advanced AI systems can automate high-skill cognitive work, including scientific research, engineering design, and policy analysis, then, in principle, human development should no longer be geographically concentrated.
If automated intelligence becomes a commodity for everyone, does your geographic and socioeconomic position still matter?

\paragraph{Promises and Tension.}

Universal access to expert-level knowledge, automated scientific discovery, democratization of healthcare, and AI-mediated governments could allow societies in the Global South to bypass traditional development bottlenecks and finally overcome historical constraints \citep{openai_planning_for_agi, deepmind_responsible_agi}.
At the same time, early empirical signals from current AI deployment suggest that access to these capabilities is heavily controlled by infrastructure ownership, institutional capabilities, and geopolitical agendas. 
Compute concentration, proprietary models, and AI-mediated allocation systems already benefit actors embedded in formal, data-rich environments. 
Succintly, privilege and power control these capabilities.
These dynamics introduce a tension at the core of the AGI promise: systems designed to generalize intelligence may amplify pre-existing asymmetries in access, control, and relevance \citep{hao2025empire}.

The following section explores how this tension and promises may resolve, outlining three plausible futures for the Global South under an AGI regime.

%% file: 3_futures.tex
\begin{figure}[t!]
\centering
  \includegraphics[width=\linewidth]{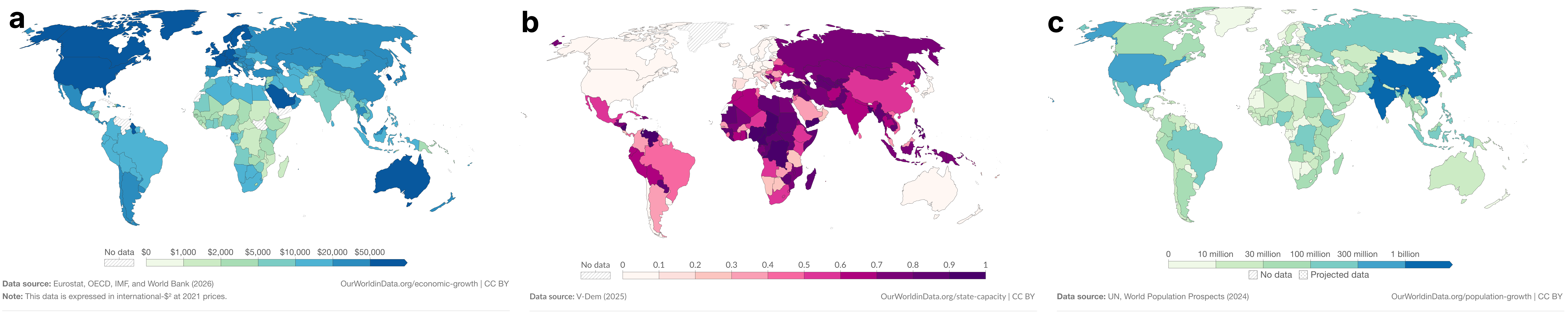}
  \caption{
    \textbf{Global Structural Asymmetries on the Impact of AGI.} 
    \textbf{a)} GDP per capita highlights the concentration of economic output outside much of the Global South. 
    \textbf{b)} Political corruption perception index illustrates institutional fragility across large regions of the Global South. 
    \textbf{c)} Global population distribution shows that the majority of the world’s population resides in regions with comparatively low economic output and weaker institutional capacity. 
    Together, these panels illustrate the coexistence of large latent human potential and persistent structural constraints, shaping whether AGI enables broad-based empowerment or selective exclusion.
    \footnotesize{Figure sources: Our World in Data~\cite{owid_gdp, owid_corruption, owid_population}.
    }}
  \label{fig:futures}
\end{figure}
\section{The Future of the Global South}

We consider that AGI will generate a clear separation in potential pathways for global outcomes, with particularly high stakes for people in the Global South. 
Existing structural inequalities may be eliminated, increases, or reconfigured in subtler but more permanent forms.

\paragraph{The Utopia: Location Stops Mattering.}
AGI systems dramatically lower barriers to knowledge, expertise, and essential services. 
Scientific discovery, education, healthcare optimization, and economic planning become globally accessible and inexpensive. 
Individuals born in geographically remote or economically marginalized regions gain access to the same diagnostic, educational, and decision-support capabilities as those in the world’s most developed urban centers.
Under this vision, access to expert-level reasoning is independent from location and wealth. 
Institutional capacity improves through AI-mediated governments, corruption and inefficiency are reduced through transparent allocation mechanisms, and social development becomes less constrained by legacy infrastructure. 
The Global South benefits not through slow industrial convergence, but through rapid cognitive jumps that bypasses historical development bottlenecks tied to socioeconomic constraints.

\paragraph{The Collapse: From Exploitation to Irrelevance.}
AGI eliminates the economic necessity of large human populations altogether. 
Unlike historical colonial systems, which depended on extracting labor and resources from peripheral regions, an AGI-dominated economy may no longer require either. 
Scientific production, capital accumulation, and political power concentrate within a small number of technologically sovereign regions, or even among a narrow set of individuals with privileged access to AGI systems.
In this setting, large portions of the Global South are not exploited, they are bypassed. 
Their labor becomes unnecessary as their markets are marginal, and their political leverage collapses. 
Remaining relevance is limited primarily to land and natural resources, which may be extracted with minimal local participation. 
This represents a qualitative shift from historical inequality: populations once integrated through extraction are rendered irrelevant to systems that no longer depend on their participation due to full or near-full automation.

\paragraph{The Middle Ground.}
The most plausible outcome lies between these extremes and closely resembles the present reality but intensified. 
AGI systems amplify productivity and wealth for already advantaged individuals and companies while leaving underlying social and economical issues largely intact. 
Access to compute, infrastructure, institutional trust, and geopolitical goals determines who benefits.
Informal economies persist, but become increasingly disconnected from AI-mediated systems of credit, insurance, logistics, and governance. 
For those outside the elite and privileged, this produces a gradual erosion of relevance: participation shrinks, opportunities narrow, and exclusion becomes normalized. 
Inequality deepens without a clear rupture, making this trajectory both stable and difficult to change.
How much worse can conditions become for those already confined to the margins of human development?

%% file: conclusion.tex
\section{Conclusion}

Historically, global inequality has been created through extraction systems that required human labor. 
Colonialism, despite all its brutality, depended on people \citep{bufacchi2017colonialism}. 
An AGI regime may be bound to alter this logic. 
When intelligence, coordination, and government can be automated at scale, the extraction of human labor may no longer be necessary.
This marks a transition from exploitation to exclusion. 
The central risk posed by increasingly capable AI systems for the Global South is therefore not widespread job replacement, but a deeper loss of relevance within global economic and political systems that no longer depend on its labor, markets, or political cooperation. 
The danger is not only in losing your income, but the conditions under which societies and individuals are considered economically and politically important at all.

Crucially, this dynamic does not imply uniform marginalization at the level of nation-states. 
Exclusion increasingly operates at the level of individuals and social classes. 
Privileged elites within the Global South, those with access to capital, infrastructure, and transnational networks, are likely to remain integrated into global AI-driven systems. 
In contrast, individuals at the lowest end of economic participation face heightened risk of exclusion, reinforcing self-sustaining feedback loops in which local inequality deepens even as elite actors transcend national constraints.

AGI, whether realized fully or approximated through advanced AI systems, acts as a reality check for existing inequalities. 
It opens a narrow space of possibility in which geographic location may no longer constrain opportunity, but more plausibly pushes the world toward a future where exclusion is gradual, normalized, and perhaps impossible to reverse. 
The implications for the Global South are therefore neither totally dystopian nor automatically beneficial.
They depend on how access, and institutional inclusion are structured in the presence of highly autonomous AI systems.

% \section{Conclusion}

% This paper has argued that the most significant risk posed by increasingly capable AI systems for the Global South is not widespread job replacement, but a deeper loss of relevance within global economic and political systems. 
% Participation is no longer guaranteed by labor or intelligence alone, but by access to infrastructure, institutional inclusion, and structural geopolitical position.

% AGI, whether realized fully or approximated through advanced AI systems, acts as a stress test for existing inequalities. 
% It introduces a fundamental change in possible futures: one in which geographic location ceases to meaningfully constrain opportunity, and another in which large segments of humanity are bypassed by systems that no longer depend on their participation. 
% The most likely outcome lies between these extremes, where exclusion is gradual, normalized, and increasingly difficult to reverse.

% The implications for the Global South are therefore neither uniformly dystopian nor automatically emancipatory. 
% Rather, they depend on how intelligence, access, and governance are structured in the presence of highly autonomous systems. 
% In this sense, AGI does not simply amplify existing inequalities, it risks redefining the conditions under which societies and individuals are considered economically and politically consequential at all.